\documentclass[12pt]{iopart}
\usepackage{iopams}
\usepackage{graphicx}
\input amssym.def \input amssym

\begin{document}

\title[]{On small proofs of the Bell-Kochen-Specker theorem for two, three and four qubits}

\author{ Michel Planat
}

\address{ Institut FEMTO-ST, CNRS, 32 Avenue de
l'Observatoire,\\ F-25044 Besan\c con, France. }
\vspace*{.1cm}

\begin{abstract} The Bell-Kochen-Specker theorem (BKS) theorem rules out realistic {\it non-contextual} theories
 by resorting to impossible assignments of rays among a selected set of maximal orthogonal bases. We investigate the geometrical structure of
 small $v-l$ BKS-proofs  involving $v$ real rays and $l$ $2n$-dimensional bases of $n$-qubits ($1< n < 5$). Specifically, we look at the parity proof $18-9$ with two qubits (A. Cabello, 1996 \cite{Cabello1996}), the parity proof $36-11$ with three qubits (M. Kernaghan \& A. Peres, 1995 \cite{Kernaghan1965}) and a newly discovered non-parity
 proof 80-21 with four qubits (that improves work of P.~K Aravind's group in 2008 \cite{Harvey2008}).
 The rays in question arise as real eigenstates shared by some maximal commuting sets (bases) of operators in the $n$-qubit Pauli group.
 One finds characteristic signatures of the distances between the bases, which carry various symmetries in their graphs.

\end{abstract}



\section{Introduction}

Contextuality is an important hallmark of quantum mechanics. In a contextual world, the measured value of an observable depends on which other mutually compatible measurements might be performed. In this line of thought, the Bell-Kochen-Specker (BKS) theorem is fundamental because it is able to rule out non-contextual hidden variable theories \cite{Kochen1967,Peres} by resorting to mathematical statements about coloring of rays located on maximal orthonormal bases in a $d$-dimensional Hilbert space ($d \ge 3$).

 A non-coloring BKS proof consists of a finite set of rays/vectors that cannot be assigned truth values ($1$ for true, $0$ for false) in such a way that (i) one member of each complete orthonormal
  basis is true and (ii) no two orthogonal (that is, mutually compatible)
 projectors are both true \cite[p. 197]{Peres}-\cite{ Aravind1998} \footnote{Throughout the paper, the word proof is not taken in the strict mathematical sense as a list of logical statements, but as a set $v-l$ of $v$ rays and $l$ maximal bases satisfying the BKS postulates/constraints.}. The smallest state-independent proofs in three dimensions are of the $31-17$ type ($31$ rays located on $17$ orthogonal triads) and the (closely related) $33-16$ type corresponding to a very symmetric
 arrangement of rays located on a cube of edge $\sqrt{2}$ \cite[fig. 7.2, p. 198]{Peres}, see also \cite{Arends2011}. The BKS theorem is intimately related to the coloring of a graph whose vertices are the rays and whose edges are the bases \cite{Cabello2011}.

 A parity proof of BKS theorem is a set of $v$ rays that form $l$ bases ($l$ odd) such that each ray occurs an even number of times over these bases. A proof of BKS theorem is ray critical (resp. basis critical) if it cannot be further simplified by deleting even a single ray (resp. a single basis), see \cite[p. 9]{Aravind2011} and \cite{Pavicic2005}\footnote{The authors of \cite{Pavicic2005} discuss the KS theorem in terms of so-called KS sets and sometimes arrive at different counts for the minimal numbers of vectors required. }. The smallest BKS proof in dimension $4$ (resp. $8$) is a parity proof and corresponds to arrangements of real states arising from the two-qubit (resp. three-qubit) Pauli group, more specifically as eigenstates of operators forming Mermin's square (\ref{Mermin1}) (resp. Mermin's pentagram (\ref{Mermin2})) \cite{Mermin1993}.
 In what follows, we shall investigate in detail the structure of the $18-9$ two-qubit proofs \cite{Cabello1996,Aravind2011}, that of the $36-11$ three-qubit proofs \cite{Kernaghan1965}, and the related small proofs. Moreover, we shall improve the earlier four-qubit $80-265$ proof \cite{Harvey2008} by simplifying it to a $80-21$ one.

Our overall goal in this paper is to gain a deeper understanding of the algebraic and geometrical structure of the minimal BKS $n$-qubit proofs. This is not a straigthforward task because there exists a plethora of quantum states appearing as eigenstates shared by the maximal commuting sets of operators in the $n$-qubit Pauli group. The total number of states is $dL$, where $d=2^n$ and $L=\prod_{i=1}^n (1+2^i)$ is the number of maximal commuting sets, see for example \cite[eq. (16)]{Planat2011}. The number of real states is found to be $L_R=\prod_{i=1}^n (2+2^i)$, corresponding to the sequence $\{4,24,240,4320,\cdots\}$ of kissing numbers in the Barnes-Wall lattice $B_n$ of dimension $2^n$. One can ultimately expect a deep relationship between $n$-qubit BKS proofs and the $B_n$'s (in the spirit of \cite{Planat2012}),  but our goal here is more modest. We shall restrict the reservoir of real states to those generated by Mermin's square ($24$ states for two qubits), Mermin's pentagram ($40$ states for three qubits) and the magic rectangle (\ref{Mermin3}) found in \cite{Harvey2008} ($80$ states for four qubits) \footnote{The BKS theorem also admits many proofs with complex rays as already shown for the two-qubit case \cite{Aravind2011new}.}.

Apart from the use of standard graph theoretical tools for characterizing the ray/base symmetries, we shall employ a useful signature of the proofs in terms of Bengtsson's distance $D_{ab}$ between two orthonormal bases $a$ and $b$ defined as \cite[eq. (2)]{Raynal2011}-\cite{Beng2007}
\begin{equation}
D_{ab}^2=1-\frac{1}{d-1}\sum_{i,j}^d \left(\left|\left\langle a_i|b_j \right\rangle\right|^2-\frac{1}{d}\right)^2.
\label{Beng}
\end{equation}

The distance (\ref{Beng}) vanishes when the bases are the same and is maximal (equal to unity) when the two bases $a$ and $b$ are mutually unbiased, $\left|\left\langle a_i|b_j \right\rangle\right|^2=1/d$, and only then. We shall see that the bases of a BKS proof employ a selected set of distances which seems to be a universal feature of the proof \footnote{Instead of a signature built from the maximal bases one can of course define a signature built from the rays involved in the proof, as in \cite{Ruuge2012}.}.

The next three sections \ref{sec2}, \ref{sec3}, and \ref{sec4} focus on two-, three- and four-qubit proofs built from the operators in the corresponding Pauli groups. We denote by $X$, $Z$ and $Y$ the Pauli spin matrices in $x$, $y$ and $z$ directions, and the tensor product is not explicit, i. e. in (\ref{Mermin1}) one denotes $Z_1=Z  \otimes I$, $Z_2=I \otimes Z$ and $ZZ=Z\otimes Z$, in (\ref{Mermin2}) one denotes $Z_1=Z \otimes I\otimes I$ and so on, with $I$ being the identity matrix of the corresponding dimension.

 The symmetries underlying the proofs and the distances between the involved bases are revealed \footnote{The notations we use are standard ones: the symbols $\times$ and $\rtimes$ mean the direct and semidirect product of groups, $S_n$ is the $n$-letter symmetric group and $D_n$ is the $2n$-element dihedral group.}. In some sense, quantum contextuality encompasses quantum complementarity by having recourse, not only to the maximal distance corresponding to mutually unbiased bases, but also to another set of distances which is a signature of the proof. Knowing the particular set of distances used in a proof of a given type, one is able to derive all proofs of the same type and their overall structure (at least for two and three qubits).


\section{The BKS parity proofs for two qubits}
\label{sec2}

The simplification of arguments in favour of a contextual view of quantum measurements started with Peres' note \cite{Peres1990} and Mermin's report \cite{Mermin1993}. Observe that in (\ref{Mermin1}), the three operators in each row and each column mutually commute and their product is the identity matrix, except for the right hand side column whose product is minus the identity matrix. There is no way of assigning multiplicative properties to the eigenvalues $\pm 1$ of the nine operators  while still keeping the same multiplicative properties for the operators \footnote{It is intriguing that such a property can be given a ring geometrical illustration by seeing Mermin's square as the projective line over the ring $\mathbb{F}_2 \times \mathbb{F}_2$ (where $\mathbb{F}_2$ is the field with two elements) and the right hand side column as the locus for pairs of units or pair of zero divisors of $R$. Ultimately, the geometry of the $15$ two-qubit operators in the Pauli group has been found to mimic the generalized quadrangle $GQ(2,2)$, see \cite{Planat2011} and references therein.}. Paraphrasing \cite{Peres1990}, the result of a measurement depends \lq\lq in a way not understood, on the choice of other quantum measurements, that may possibly be performed".

\begin{equation}
\begin{array}{ccc}
| & | & || \\
-Z_1- & Z_2- & ZZ- \\
| & | & || \\
-X_2- & X_1- & XX- \\
| & | & ||\\
-ZX -& XZ- & YY- \\
| & | & || \\
\end{array}
\label{Mermin1}
\end{equation}

The next step to be able to see behind the scene, and to reveal the simplest paradoxical/contextual set of rays and bases, was achieved by A. Cabello \cite{Cabello1996}. It is a $18-9$ BKS parity proof that can be given a remarkable diagrammatic illustration fitting the structure of a $24$-cell \cite{Aravind2011}. More generally, it is already known \cite{Aravind2011,Pavicic2005} that there exist four main types of parity proofs arising from $24$ Peres rays \cite{Peres}, that are of the type $18-9$, $20-11$, $22-13$ and $24-15$. Types $20-11$ and $22-13$ subdivide into two non-isomorphic ones $A$ and $B$ as shown in Table 1.

For the list of the unnormalized eigenvectors (numbered consecutively) we use the same notation as \cite{Aravind2011}
\begin{eqnarray}
&1:[1 0 0 0],~~2:[0 1 0 0],~~3:[0 0 1 0],~~4:[0 0 0 1],~~5:[1 1 1 1],~~6:[ 1  1 \bar{1} \bar{1}] \nonumber\\
&7:[ 1 \bar{1}  1 \bar{1}],~~8:[ 1 \bar{1} \bar{1}  1],~~9:[ 1 \bar{1} \bar{1} \bar{1}],~~10:[ 1 \bar{1}  1  1],~~11:[ 1  1 \bar{1}  1],~~12:[ 1  1  1 \bar{1}]\nonumber\\
&13:[1 1 0 0],~~14:[ 1 \bar{1}  0  0],~~15:[0 0 1 1],~~16:[ 0  0  1 \bar{1}],~~17:[0 1 0 1],~~18:[ 0  1  0 \bar{1}]\nonumber\\
&19;[1 0 1 0],~~20:[ 1  0 \bar{1}  0],~~21:[ 1  0  0 \bar{1}],~~22:[1 0 0 1],~~23:[ 0  1 \bar{1}  0],~~24:[0 1 1 0]\nonumber\\
\label{rays24}
\end{eqnarray}

The $24$ complete orthogonal bases are as follows

\begin{eqnarray}
&1 :\{ 1, 2, 3, 4 \},
2 :\{ 5, 6, 7, 8 \},
3 :\{ 9, 10, 11, 12 \},
4 :\{ 13, 14, 15, 16 \},\nonumber\\
&5 :\{ 17, 18, 19, 20 \},
6 :\{ 21, 22, 23, 24 \},
7 :\{ 1, 2, 15, 16 \},
8 :\{ 1, 3, 17, 18 \},\nonumber\\
&9 :\{ 1, 4, 23, 24 \},
10 :\{ 2, 3, 21, 22 \},
11 :\{ 2, 4, 19, 20 \},
12 :\{ 3, 4, 13, 14 \},\nonumber\\
&13 :\{ 5, 6, 14, 16 \},
14 :\{ 5, 7, 18, 20 \},
15 :\{ 5, 8, 21, 23 \},
16 :\{ 6, 7, 22, 24 \},\nonumber\\
&17 :\{ 6, 8, 17, 19 \},
18 :\{ 7, 8, 13, 15 \},
19 :\{ 9, 10, 13, 16 \},
20 :\{ 9, 11, 18, 19 \},\nonumber\\
&21 :\{ 9, 12, 22, 23 \},
22 :\{ 10, 11, 21, 24 \},
23 :\{ 10, 12, 17, 20 \},
24 :\{ 11, 12, 14, 15 \}\nonumber\\
\end{eqnarray}

Normalizing rays (\ref{rays24}), a finite set of distances (\ref{Beng}) between the $24$ bases is found to be
$$D=\{a_1,a_2,a_3,a_4,a_5\}=\{\frac{1}{\sqrt{3}},\frac{\sqrt{7}}{\sqrt{12}},\frac{\sqrt{2}}{\sqrt{3}},\frac{\sqrt{5}}{\sqrt{6}},1\}\approx \{0.577,0.763,0.816,0.912,1.000\}.$$
 Table \ref{table1} provides a histogram of distances for various parity proofs $v-l$.
\begin{table}[ht]
\begin{center}
\small
\begin{tabular}{|r||r||r|r|r|r|r|}
\hline
 proof $v-l$ & \# proofs & $a_1$ &  $a_2$ &$a_3$ &  $a_4$ &  $a_5$\\
\hline
$24-15$ & $16$ & $18$ &   $18$ & $9$ &   $54$&   $6$\\
\hline
$22-13A$ &$96$ &  $12$ &   $18$ & $3$ &   $42$&   $3$\\
$22-13B$ &$144$ &  $12$ &   $18$ & $4$ &   $42$&   $2$\\
\hline
$20-11A$ & $96$ & $6$ &   $18$ & $0$ &   $30$&   $1$\\
$20-11B$ & $144$ & $6$ &   $18$ & $1$ &   $30$&   $0$\\
\hline
$18-9$  & $16$ & $0$  &   $18$ & $0$ &   $18$&   $0$\\
\hline
\end{tabular}
\label{table1}
\caption{The histogram of distances for various parity proofs $v-l$ obtained from Mermin's square. One can check the expected equality $2\sum a_i=l(l-1)$ in each proof.
Let us first observe that the symmetry group of Mermin's graph (\ref{Mermin1}) is $G_{72}=\mathbb{Z}_3^2 \rtimes D_4.$
The $16$ proofs of the $18-9$ type overlap in $3$ or $5$ elements. The way the proofs overlap each other (the crossing graph) is that of the square (\ref{MagicSquare})
with $\mbox{aut}\cong \mathbb{Z}_2^4 \rtimes G_{72}$. For the $16$ proofs of the $24-15$ type, the symmetry is the same.
Basically, still the same group governs the $240=96+144$ proofs of the $20-11$ type (as well as the $240$ proofs of the $22-13$ type), although there also exist some extra abelian symmetries. (For a nice geometrical display of the proofs, see \cite{Aravind2011}).
}
\end{center}
\end{table}

Tables 1 and 2 give all essential information about the proofs. First, a proof of a given type possesses a seemingly universal pattern in terms of the distances. Observe that the smallest proof does not contain any pair of mutually unbiased bases. Second, a given proof can be seen from its symmetry subsets, each one attached  to a selected crossing graph (see the captions of Tables 1 and 2). Then, one can create a graph having the bases as vertices and an edge joining two vertices if the two bases are in the proper range of distances. The cliques of the latter graph (not all maximal), of the selected odd size $l$, are candidates for a proof of the $v-l$ type, but not all of them provide proofs.   This way, we could explicit all the proofs, $16$ proofs of the $18-9$ type (as for the $24-15$ type) and $240$ proofs of the $20-11$ type (as for the $22-13$ type), as reported, for example, in \cite[Table 2]{Aravind2011}.

\begin{table}[ht]
\begin{center}
\small
\begin{tabular}{|r||r|r|r|r|}
\hline
 \# common elements & $0$ &  $1$ & $2$  \\
\hline
 distance &$a_5,a_4$&$a_2$&$a_1$\\
\hline
\hline
$24-15$ type & $G_{72}$ &   $\mathbb{Z}_2^9 \rtimes (D_4\times S_6)$ & $G_{72}$ \\
\hline
$22-13A$ type &   $D_4$ &   $G_{72}\times S_4$ & $D_4$\\
$22-13B$ type &   $D_4$ &   $\mathbb{Z}_3^2 \rtimes (\mathbb{Z}_2^4 \rtimes D_6)$ & $D_4$\\
\hline
$20-11A$ type &    $D_6$ &   $\mathbb{Z}_3^2 \rtimes \mathcal{P}_1$ & $\mathbb{Z}_3^3 \rtimes \mathcal{P}_1$ \\
$20-11B$ type &    $D_4$ &   $\mathbb{Z}_3^2 \rtimes \mathcal{P}_1$ & $\mathbb{Z}_2^4 \rtimes D_6$ \\

\hline
$18-9$ type &   $G_{72}$ &   $G_{72}$ & no overlap \\
\hline
\end{tabular}
\label{table2}
\caption{The symmetries involved in various two-qubit parity proofs of the BKS theorem. The first row gives the number of common elements between the bases, the second row relates these numbers to distinguished distances. Among the building block symmetries are the group $G_{72}=\mathbb{Z}_3^2 \rtimes D_4$ and the single qubit Pauli group $\mathcal{P}_1 \cong D_4 \rtimes \mathbb{Z}_2$, a group underlying the CPT symmetries of the Dirac equation \cite{PlanatCPT}. }
\end{center}
\end{table}

The $16$ proofs of the $18-9$ type can be displayed as the $4 \times 4$ square (\ref{MagicSquare}) in which two adjacent proofs share three bases. Observe that each $2 \times 2$ square of adjacent proofs has the same shared base, which is taken
as an index (e.g. the upper left-hand-side $2 \times 2$ square has index $7$ and the lower right-hand-side square has index $10$). All four indices in each row and in each column correspond to four disjoint bases that together partition the $24$ rays.

\footnotesize
\begin{equation}
\begin{array}{cccc}
 \left(\begin{array}{ccc} 7 &8  &10 \\13 & 14  &16 \\22 &23 &24  \end{array}\right)- &\left(\begin{array}{ccc} 7 &9  &11 \\14 & 15  &18 \\19 &20 &22 \end{array}\right)-
&\left(\begin{array}{ccc} 8 &9  &12 \\16 & 17  &18 \\20 &21 &24 \end{array}\right)- &\left(\begin{array}{ccc} 10 &11  &12 \\13 & 15  &17 \\19 &21 &23 \end{array}\right)-\\
|_7&|~_{20}&|_{12}&|_{23}\\
\left(\begin{array}{ccc} 7 &9  &11 \\16 & 17  &18 \\19 &21 &23 \end{array}\right)- &\left(\begin{array}{ccc} 7 &8  &10 \\13 & 15  &17 \\20 &21 &24 \end{array}\right)-
&\left(\begin{array}{ccc} 10 &11  &12 \\13 & 14  &16 \\19 &20 &22 \end{array}\right)- &\left(\begin{array}{ccc} 8 &9  &12 \\14 & 15  &18 \\22 &23 &24  \end{array}\right)-\\
|_{17}&|_{10}&|_{14}&|_9\\
\left(\begin{array}{ccc} 8 &9  &12 \\13 & 15  &17 \\19 &20 &22 \end{array}\right)- &\left(\begin{array}{ccc} 10 &11  &12 \\16 & 17  &18 \\22 &23 &24 \end{array}\right)-
&\left(\begin{array}{ccc} 7 &8  &10 \\14 & 15  &18 \\19 &21 &23 \end{array}\right)- &\left(\begin{array}{ccc} 7 &9  &11 \\13 & 14  &16 \\20 &21 &24  \end{array}\right)-\\
|_{12}&|_{23}&|_7&|_{20}\\
\left(\begin{array}{ccc} 10 &11 &12 \\14 & 15  &18 \\20 &21 &24 \end{array}\right)- &\left(\begin{array}{ccc} 8 &9  &12 \\13 & 14  &16 \\19 &21 &23 \end{array}\right)-
&\left(\begin{array}{ccc} 7 &9  &11 \\13 & 15  &17 \\22 &23 &24 \end{array}\right)- &\left(\begin{array}{ccc} 7 &8  &10 \\16 & 17  &18 \\19 &20 &22  \end{array}\right)-\\
|_{14}&|_9&|_{17}&|_{10}\\
 \end{array}
\label{MagicSquare}
\end{equation}
\normalsize

The $16$ proofs of the $24-15$ type (not shown) also form a $4 \times 4$ square
 in which two proofs share seven elements, comprising a common part of the six reference bases $1-6$  and an isolated base.

\subsection*{Diagrams for the proofs}

How can we account for the distance signature of a given proof? A simple diagram does the job.

The diagram for the $18-9$ proof is simply a $3 \times 3$ square. Below we give an explicit construction of the first proof that corresponds to the upper left-hand-side corner in (\ref{MagicSquare}).
 The $9$ vertices of the graph are the $9$ bases of the proof, the one-point crossing graph between the bases is the graph (\ref{MagicSquare9}), with $\mbox{aut}=G_{72}=\mathbb{Z}_3^2 \rtimes D_4$.
There are $18$ (distinct) edges that encode the $18$ rays, a selected vertex/base of the graph is encoded by the union of the four edges/rays that are adjacent to it.

\begin{equation}
\begin{array}{ccc}
 \left(\begin{array}{cc} 1 &2  \\15 & 16    \end{array}\right)- 1-&\left(\begin{array}{cc} 1 &3  \\17 & 18  \end{array}\right)-3-
&\left(\begin{array}{cc} 2 &3   \\21 & 22   \end{array}\right)- 2\\
|_{16}&|~_{18}&|_{22}\\
\left(\begin{array}{cc} 5 &6   \\14 & 16  \end{array}\right)-5- &\left(\begin{array}{cc} 5 &7  \\18 & 20   \end{array}\right)-7-
&\left(\begin{array}{cc} 6 &7   \\21 & 24    \end{array}\right)-6 \\
|_{14}&|_{20}&|_{24}\\
\left(\begin{array}{cc} 11 &12   \\14 & 15   \end{array}\right)-12- &\left(\begin{array}{cc} 10 &12   \\17 & 20 \end{array}\right)-10-
&\left(\begin{array}{cc} 10 &11   \\21 & 24   \end{array}\right)-11 \\
|_{15}&|_{17}&|_{21}\\
 \end{array}
\label{MagicSquare9}
\end{equation}

As for the distances between the bases, two bases located in the same row (or the same column) have distance $a_2=\sqrt{7/12}$, while two bases not in the same row (or column) have distance $a_4=\sqrt{5/6}>a_2$, as readily discernible from Table 2 and the histogram in Table 1. Indeed, any  proof of the $18-9$ type has the same diagram as (\ref{MagicSquare9}).

Similar diagrams can be drawn to reflect the histogram of distances in proofs of a larger size. Below we restrict to the case of  a $20-11A$ proof (where only the distance between two bases is made explicit, but not the common rays of the bases)

\small
\begin{equation}
\begin{array}{cccc}
 \left(\begin{array}{cc} 10 &12  \\17 & 20    \end{array}\right)- a_2-&\left(\begin{array}{cc} 11 &12  \\14 & 15  \end{array}\right)-a_2-
&\left(\begin{array}{cc} 10 &11  \\21 & 24   \end{array}\right)... a_4=\sqrt{5/6}... &\\
|{a_2=\sqrt{7/12}}&|~a_2&|a_2& ...\\
\left(\begin{array}{cc} 1 &3   \\17 & 18  \end{array}\right)-a_2- &\left(\begin{array}{cc} 1&2  \\15 & 16   \end{array}\right)-a_2-
&\left(\begin{array}{cc} 1 &4   \\23 & 24    \end{array}\right)..  a_1=\frac{1}{\sqrt{3}}..& \left(\begin{array}{cc} 1 &2  \\3 & 4  \end{array}\right)\\
|a_2&|a_2&|a_2&  |a_5=1\\
\left(\begin{array}{cc} 5 &7   \\18 & 20   \end{array}\right)-a_2- &\left(\begin{array}{cc} 5 &6   \\14 & 16 \end{array}\right)-a_2-
&\left(\begin{array}{cc} 5 &8   \\21 & 23   \end{array}\right).. a_1=\frac{1}{\sqrt{3}}..&\left(\begin{array}{cc} 5 &6  \\7 & 8  \end{array}\right)\\
|a_2&|a_2&|a_2&...\\
 \end{array}
\label{MagicSquare11}
\end{equation}
\normalsize

The proof consists of $11$ bases, $9$ of them have the same mutual diagram as in (\ref{MagicSquare9}) and their mutual distance is $a_2=\sqrt{7/12}$ (as shown) or $a_4=\sqrt{5/6}$ (not shown), depending on whether they are located in the same row (or the same column) of the $3 \times 3$ square, or not. The extra two bases of the right-hand-side column are mutually unbiased (with distance $a_5=1$), their distance to any base of the same row is $1/\sqrt{3}$ and their distance to any base of the first row is $a_4$ (as shown).


\section{The BKS parity proofs for three qubits}
\label{sec3}

Quantum contextuality of a three-qubit system is also predicted in Mermin's report \cite{Mermin1993} in terms of its famous pentagram. Below we display it in a sligthly different form in order to underline its kinship to the four-qubit \lq\lq magic" rectangle (\ref{Mermin3}). Mermin's rectangle/pentagram (\ref{Mermin2}) features the same (real) operators as in \cite{Saniga2012} \footnote{In \cite{Saniga2012}, it is shown that Mermin's pentagram corresponds to an ovoid of the three-dimensional projective space of order two, $PG(3,2)$, which generalizes the results discussed in the footnote on p. 3.}.

\begin{equation}
\begin{array}{cccc}
| & | & | & |\\
Z_1 & Z_1 & X_1 & X_1\\
| & | & | & |\\
Z_2 & X_2 & Z_2 & X_2\\
| & | & | & |\\
Z_3 & X_3 & X_3 & Z_3 \\
| & | & | & |\\
=ZZZ =& ZXX =&XZX=& XXZ =\\
| & | & | & |\\
\end{array}
\label{Mermin2}
\end{equation}

Following \cite{Mermin1993}, (\ref{Mermin2}) is a parity proof of the BKS theorem because mutually commuting operators in the four columns multiply to the identity matrix while operators in the single row multiply to minus the identity matrix. Since each operator appears twice in this reasoning, it is impossible to assign truth values $\pm 1$ to the eigenvalues while keeping the multiplicative properties of the operators.

The list of (unormalized) eigenvectors coming from the five bases in (\ref{Mermin2}) is (in the notation of \cite{Kernaghan1965})

\small
\begin{eqnarray}
&1
:[1 0 0 0 0 0 0 0],
2
:[0 1 0 0 0 0 0 0],
3
:[0 0 1 0 0 0 0 0],
4
:[0 0 0 1 0 0 0 0],
5
:[0 0 0 0 1 0 0 0],\nonumber\\
&6
:[0 0 0 0 0 1 0 0],
7
:[0 0 0 0 0 0 1 0],
8
:[0 0 0 0 0 0 0 1],
9
:[1 1 1 1 0 0 0 0],
10
:[ 1  1 \bar{1} \bar{1}  0  0  0  0],\nonumber\\
&11
:[ 1 \bar{1}  1 \bar{1}  0  0  0  0],
12
:[ 1 \bar{1} \bar{1}  1  0  0  0  0],
13
:[0 0 0 0 1 1 1 1],
14
:[ 0  0  0  0  1  1 \bar{1} \bar{1}],
15
:[ 0  0  0  0  1 \bar{1}  1 \bar{1}],\nonumber\\
16
&:[ 0  0  0  0  1 \bar{1} \bar{1}  1],
17
:[1 1 0 0 1 1 0 0],
18
:[ 1  1  0  0 \bar{1} \bar{1}  0  0],
19
:[ 1 \bar{1}  0  0  1 \bar{1}  0  0],
20
:[ 1 \bar{1}  0  0 \bar{1}  1  0  0],\nonumber\\
&21
:[0 0 1 1 0 0 1 1],
22
:[ 0  0  1  1  0  0 \bar{1} \bar{1}],
23
:[ 0  0  1 \bar{1}  0  0  1 \bar{1}],
24
:[ 0  0  1 \bar{1}  0  0 \bar{1}  1],
25
:[1 0 1 0 1 0 1 0],\nonumber\\
&26
:[ 1  0  1  0 \bar{1}  0 \bar{1}  0],
27
:[ 1  0 \bar{1}  0  1  0 \bar{1}  0],
28
:[ 1  0 \bar{1}  0 \bar{1}  0  1  0],
29
:[0 1 0 1 0 1 0 1],
30
:[ 0  1  0  1  0 \bar{1}  0 \bar{1}],\nonumber\\
&31
:[ 0  1  0 \bar{1}  0  1  0 \bar{1}],
32
:[ 0  1  0 \bar{1}  0 \bar{1}  0  1],
33
:[ 1  0  0  1  0  1 \bar{1}  0],
34
:[ 1  0  0 \bar{1}  0  1  1  0],
35
:[ 1  0  0  1  0 \bar{1}  1  0],\nonumber\\
36
&:[ 1  0  0 \bar{1}  0 \bar{1} \bar{1}  0],
37
:[ 0  1  1  0 \bar{1}  0  0  1],
38
:[ 0  1 \bar{1}  0  1  0  0  1],
39
:[ 0  1 \bar{1}  0 \bar{1}  0  0 \bar{1}],
40
:[ 0  1  1  0  1  0  0 \bar{1}].\nonumber\\
\label{40rays}
\end{eqnarray}
\normalsize

These rays form $25$ maximal orthogonal bases

\footnotesize
\begin{eqnarray}
&1 :\{ 1, 2, 3, 4, 5, 6, 7, 8 \},
2 :\{ 1, 2, 3, 4, 13, 14, 15, 16 \},
3 :\{ 1, 2, 5, 6, 21, 22, 23, 24 \},\nonumber\\
&4 :\{ 1, 3, 5, 7, 29, 30, 31, 32 \},
5 :\{ 1, 4, 6, 7, 37, 38, 39, 40 \},
6 :\{ 5, 6, 7, 8, 9, 10, 11, 12 \},\nonumber\\
&7 :\{ 9, 10, 11, 12, 13, 14, 15, 16 \},
8 :\{ 9, 10, 13, 14, 19, 20, 23, 24 \},
9 :\{ 9, 11, 13, 15, 27, 28, 31, 32 \},\nonumber\\
&10 :\{ 9, 12, 14, 15, 34, 36, 38, 39 \},
11 :\{ 10, 11, 13, 16, 33, 35, 37, 40 \},
12 :\{ 10, 12, 14, 16, 25, 26, 29, 30 \},\nonumber\\
&13 :\{ 11, 12, 15, 16, 17, 18, 21, 22 \},
14 :\{ 3, 4, 7, 8, 17, 18, 19, 20 \},
15 :\{ 17, 18, 19, 20, 21, 22, 23, 24 \},\nonumber\\
&16 :\{ 17, 19, 21, 23, 26, 28, 30, 32 \},
17 :\{ 17, 20, 22, 23, 35, 36, 37, 39 \},
18 :\{ 18, 19, 21, 24, 33, 34, 38, 40 \},\nonumber\\
&19 :\{ 18, 20, 22, 24, 25, 27, 29, 31 \},
20 :\{ 2, 4, 6, 8, 25, 26, 27, 28 \},
21 :\{ 25, 26, 27, 28, 29, 30, 31, 32 \},\nonumber\\
&22 :\{ 25, 28, 30, 31, 33, 36, 37, 38 \},
23 :\{ 26, 27, 29, 32, 34, 35, 39, 40 \},\nonumber\\
&24 :\{ 2, 3, 5, 8, 33, 34, 35, 36 \},
25 :\{ 33, 34, 35, 36, 37, 38, 39, 40 \}.\nonumber\\
\label{25bases}
\end{eqnarray}
\normalsize

\begin{table}[ht]
\begin{center}
\small
\begin{tabular}{|r||r||r|r|r|r|r|}
\hline
 proof $v-l$& \# proofs &$a_1$ &  $a_2$ &$a_3$ \\
\hline
\hline
$40-15$ & $64$ & $20$  &   $30$ & $55$ \\
\hline
$38-13$ & $640$ & $12$  &   $30$ & $26$ \\
\hline
\hline
$36-11$ & $320$ & $4$  &   $30$ & $21$ \\
\hline
\end{tabular}
\label{table3}
\caption{The histogram of distances for various parity proofs $v-l$ obtained from Mermin's pentagram.
Observe that the symmetry group of Mermin's pentagram is $S_5$. Two proofs of the $36-11$ type share $3$, $4$, $7$, $8$ or $9$ elements, with crossing graph whose $\mbox{aut}\cong \mathbb{Z}_2^6 \rtimes S_5$, or $5$ or $6$ elements with crossing graph having $\mbox{aut}\cong \mathbb{Z}_2^{14} \rtimes ( \mathbb{Z}_2 \times S_5)$.
Two proofs of the $40-15$ type have $9$, $10$, $11$ or $12$ elements in common. The graphs corresponding to $9$ or $11$ shared elements are complementary, with $\mbox{aut}\cong \mathbb{Z}_2^{10} \rtimes(A_6^2 \rtimes D_4)$,
the graph corresponding to $10$ shared elements has $\mbox{aut}\cong \mathbb{Z}_2^{32} \rtimes(\mathbb{Z}_2^5 \rtimes S_6)$ and the graph corresponding to  $12$ common elements has $\mbox{aut}\cong \mathbb{Z}_2^6 \rtimes S_5$.
}
\end{center}
\end{table}

\begin{table}[ht]
\begin{center}
\small
\begin{tabular}{|r||r|r|r|r|r|}
\hline
 \# common elements & $0$ &  $2$ & $4$  \\
\hline
 distance &$a_3$&$a_2$&$a_1$\\
\hline
\hline
$40-15$ type & $S_5$ &   $S_5^2$ & $S_5$\\
\hline
$38-13$ type &   $D_6$ &   $\mathbb{Z}_3 \rtimes (\mathbb{Z}_2 \times S_5)$ & $\mathbb{Z}_2^3 \rtimes \mathbb{Z}_6$ \\
\hline
$36-11$ type &    $\mathbb{Z}_2^2 \rtimes S_6$ &   $S_5$ & $\mathbb{Z}_2^2 \rtimes (\mathbb{Z}_6 \times S_6)$ \\
\hline
\end{tabular}
\label{table4}
\caption{The symmetries involved in various three-qubit parity proofs of the BKS theorem. The first row gives the number of common elements between the bases, this number being related in the second row to the distance between the bases. The five-letter symmetric group $S_5$ is an important building block symmetry of the proofs.}
\end{center}
\end{table}

The finite set of distances involved is
$$D=\{\frac{\sqrt{3}}{\sqrt{7}},\frac{\sqrt{9}}{\sqrt{14}},\frac{\sqrt{6}}{\sqrt{7}}\}\approx\{0.654,~0.801~,0.925\}.$$
 It contrast to the two-qubit case, there is no set of mutually unbiased bases.
Three types of parity proofs may be found, the $36-11$ type discovered in \cite{Kernaghan1965} and the two extra types $38-13$ and $40-15$. The same result (and much more) is found in \cite{WaegAra3QB}.

Tables 3 and 4 gather the main properties. As in the two-qubit case, one uses computer
to construct a graph having the bases as vertices and an edge joining two vertices/bases at the proper distances. Then one extracts all sets of cliques, not necessarily maximal, of a given odd cardinality (that is eleven, thirteen and fifteen) and keeps those having the desired property of being parity proofs of the BKS theorem. Doing this, one gets an explicit list of $64$ proofs of the $40-15$ type, $640$ proofs of the $38-13$ type and 320 proofs of the $36-11$ type, totalling to $2^{10}$ distinct parity proofs.

Below, we provide a short list of $36-11$ proofs: the $16$ proofs containing bases $1$, $2$ and $3$

\begin{eqnarray}
&1 :\{ 1, 2, 3, 4, 8, 9, 11, 16, 17, 23, 24 \},
2 :\{ 1, 2, 3, 4, 8, 9, 11, 18, 19, 22, 24 \},\nonumber\\
&3 :\{ 1, 2, 3, 4, 8, 10, 12, 16, 17, 22, 24 \},
4 :\{ 1, 2, 3, 4, 8, 10, 12, 18, 19, 23, 24 \},\nonumber\\
&5 :\{ 1, 2, 3, 4, 9, 10, 13, 16, 18, 23, 24 \},
6 :\{ 1, 2, 3, 4, 9, 10, 13, 17, 19, 22, 24 \},\nonumber\\
&7 :\{ 1, 2, 3, 4, 11, 12, 13, 16, 18, 22, 24 \},
8 :\{ 1, 2, 3, 4, 11, 12, 13, 17, 19, 23, 24 \},\nonumber\\
&9 :\{ 1, 2, 3, 5, 8, 9, 11, 16, 17, 20, 22 \},
10 :\{ 1, 2, 3, 5, 8, 9, 11, 18, 19, 20, 23 \},\nonumber\\
&11 :\{ 1, 2, 3, 5, 8, 10, 12, 16, 17, 20, 23 \},
12 :\{ 1, 2, 3, 5, 8, 10, 12, 18, 19, 20, 22 \},\nonumber\\
&13 :\{ 1, 2, 3, 5, 9, 10, 13, 16, 18, 20, 22 \},
14 :\{ 1, 2, 3, 5, 9, 10, 13, 17, 19, 20, 23 \},\nonumber\\
&15 :\{ 1, 2, 3, 5, 11, 12, 13, 16, 18, 20, 23 \},
16 :\{ 1, 2, 3, 5, 11, 12, 13, 17, 19, 20, 22 \}.\nonumber\\
\label{proofsbis}
\end{eqnarray}

These $16$ selected proofs have $4$, $5$, $6$, $7$ or $8$ bases in common. The $8$-base crossing graph is regular, of valency $5$,
with automorphism group $\mbox{aut}=\mathbb{Z}_2^5 \rtimes G_{72}$, where $G_{72}=\mathbb{Z}_3^2 \rtimes D_4$ was already found as an important
symmetry group of the two-qubit $36-11$ proofs.

\subsection*{Diagrams for the proofs}

To be more explicit, the first parity proof in (\ref{proofsbis}) consists of the eleven $8$-ray bases (\ref{eleven}), where the four rays $12$, $18$, $25$ and $38$ do not appear and the remaining ones occur $2$ or $4$ times each

\small
\begin{eqnarray}
&1:\{ 1, 2, 3, 4, 5, 6, 7, 8 \},~
2:\{ 10, 11, 13, 16, 33, 35, 37, 40\},\nonumber\\
&3:\{17, 19, 21, 23, 26, 28, 30, 32  \},~
4:\{17, 20, 22, 23, 35, 36, 37, 39 \},\nonumber\\
&5:\{ 9, 11, 13, 15, 27, 28, 31, 32 \},~
6:\{ 1, 3, 5, 7, 29, 30, 31, 32\},\nonumber\\
&7:\{ 2, 3, 5, 8, 33, 34, 35, 36 \},~
8:\{ 1, 2, 3, 4, 13, 14, 15, 16 \},\nonumber\\
&9:\{  1, 2, 5, 6, 21, 22, 23, 24  \},~
10:\{ 26, 27, 29, 32, 34, 35, 39, 40 \},\nonumber\\
&11:\{ 9, 10, 13, 14, 19, 20, 23, 24 \}.\nonumber\\
\label{eleven}
\end{eqnarray}
\normalsize

As in the previous section, a simple diagram illustrates how distances between the bases are distributed. Let us look at the $36-11$ parity proof (\ref{eleven}). The $11$ bases are displayed
as a pentagram (\ref{Pentagram}) plus the isolated reference base $1$.

\begin{equation}
\begin{array}{ccccccccc}
&&&&2&&&&\\
6 &-& &7 &- &8& &-&9  ~~==~~1 \\
&&10&&&&11&&\\
&&&&3&&&&\\
&4&&&&&&5&\\
\end{array}
\label{Pentagram}
\end{equation}

 Two adjacent bases of the pentagram have two rays in common. The reference base has with each of the bases on the horizontal line of the pentagram
 four rays in common and is disjoint from any other base. One can further observe that each line of the pentagram shares a set of four rays that is disjoint from the set of four rays shared
by another line. The automorphism group of this configuration is isomorphic to $S_5$.

 The maximal distance, $a_3$, is that between two disjoint bases, and amounts to $\sqrt{6/7}$. The intermediate distance,
 $a_2=\sqrt{9/14}$, occurs between two bases located in any line of the pentagram. Finally, the shortest distance, $a_1=\sqrt{3/7}$, is that between the reference base and each of the four bases on the horizontal line of the pentagram. Similar diagrams can be produced for any proof.

\section{The BKS proofs for four qubits}
\label{sec4}

The BKS theorem for four qubits was investigated in \cite{Harvey2008}. The \lq\lq magic" rectangle (\ref{Mermin3}) (also shown in a pentagram form in (\ref{Mermin3bis})) is a parity proof similar to (\ref{Mermin1}) and (\ref{Mermin2}) because each operator appears twice, the mutually commuting operators in any column multiply to give the identity operator and the operators in the single row multiply to give minus the identity operator. There is no way of assigning the eigenvalues $\pm 1$ while still preserving the multiplicative properties of the operators \footnote{For a finite geometrical account of the \lq\lq magic" rectangle (\ref{Mermin3}), see \cite{SanPla2012}.}.

\begin{equation}
\begin{array}{cccc}
| & | & | & |\\
Z_1 & Z_1 & X_1 & X_1\\
| & | & | & |\\
X_2 & X_2 & X_2 & X_2\\
| & | & | & |\\
Z_3 & X_3 & Z_3 & X_3 \\
| & | & | & |\\
X_4 & Z_4 & Z_4 & X_4 \\
| & | & | & |\\
=ZXZX =& ZXXZ =&XXZZ=& XXXX =\\
| & | & | & |\\
\end{array}
\label{Mermin3}
\end{equation}

\vspace{1CM}

\begin{equation}
\begin{array}{ccccccccc}
&&&&X_3&&&&\\
XXZZ && &XXXX & &ZXXZ& &&ZXZX \\
&&X_1&&&&Z_1&&\\
&&&&Z_3&&&&\\
&X_2,X_4&&&&&&X_2,Z_4&\\
\end{array}
\label{Mermin3bis}
\end{equation}

To investigate a state proof of the BKS theorem, we have at our disposal the following set of $5 \times 16=80$ rays (\ref{80rays}) and the corresponding $625$ maximal orthogonal bases

\begin{eqnarray}
&1:[1 0 1 0 1 0 1 0 0 0 0 0 0 0 0 0],2:[0 0 0 0 0 0 0 0 1 0 1 0 1 0 1 0],3:[ 0  0  0  0  0  0  0  0  1  0  1  0 \bar{1}  0 \bar{1}  0],\nonumber \\
&4:[ 0  0  0  0  0  0  0  0  1  0 \bar{1}  0  1  0 \bar{1}  0],5:[ 0  1  0 \bar{1}  0 \bar{1}  0  1  0  0  0  0  0  0  0  0],6:[ 0  1  0 \bar{1}  0  1  0 \bar{1}  0  0  0  0  0  0  0  0],\nonumber \\
&7:[ 1  0  1  0 \bar{1}  0 \bar{1}  0  0  0  0  0  0  0  0  0],8:[ 0  0  0  0  0  0  0  0  1  0 \bar{1}  0 \bar{1}  0  1  0],\nonumber \\
& \cdots \mbox{partners}\nonumber \\
%
&17:[0 0 1 0 0 0 1 0 0 0 1 0 0 0 1 0],18:[1 0 0 0 1 0 0 0 1 0 0 0 1 0 0 0],19:[ 0  1  0  0  0 \bar{1}  0  0  0  1  0  0  0 \bar{1}  0  0],\nonumber \\
&20:[ 0  0  0  1  0  0  0 \bar{1}  0  0  0  1  0  0  0 \bar{1}],21:[ 0  1  0  0  0  1  0  0  0 \bar{1}  0  0  0 \bar{1}  0  0],22:[ 0  0  0  1  0  0  0  1  0  0  0 \bar{1}  0  0  0 \bar{1}],\nonumber \\
&23:[ 1  0  0  0 \bar{1}  0  0  0 \bar{1}  0  0  0  1  0  0  0],24:[ 0  1  0  0  0 \bar{1}  0  0  0 \bar{1}  0  0  0  1  0  0],\nonumber \\
& \cdots \mbox{partners}\nonumber \\
&33:[0 0 0 0 0 0 0 0 0 0 1 1 0 0 1 1],34:[0 0 0 0 0 0 0 0 1 1 0 0 1 1 0 0],35:[ 0  0  0  0  0  0  0  0  0  0  1 \bar{1}  0  0 \bar{1}  1],\nonumber \\
&36:[ 0  0  0  0  0  0  0  0  1 \bar{1}  0  0  1 \bar{1}  0  0],37:[ 0  0  0  0  0  0  0  0  1  1  0  0 \bar{1} \bar{1}  0  0],38:[1 \bar{1}  0  0  1 \bar{1}  0  0  0  0  0  0  0  0  0  0],\nonumber \\
&39:[ 0  0  0  0  0  0  0  0  0  0  1 1 0 0  \bar{1} \bar{1}],40:[ 0  0  1 \bar{1}  0  0 \bar{1}  1  0  0  0  0  0  0  0  0],\nonumber \\
& \cdots \mbox{partners}\nonumber \\
%
&49:[ 1 \bar{1}  1 \bar{1} \bar{1}  1 \bar{1}  1  1 \bar{1}  1 \bar{1} \bar{1}  1 \bar{1}  1],50:[ 1  1  1  1 \bar{1} \bar{1} \bar{1} \bar{1} \bar{1} \bar{1} \bar{1} \bar{1}  1  1  1  1],51:
[ 1  1  1  1  1  1  1  1 \bar{1} \bar{1} \bar{1} \bar{1} \bar{1} \bar{1} \bar{1} \bar{1}],\nonumber \\
&52:[ 1  1 \bar{1} \bar{1}  1  1 \bar{1} \bar{1} \bar{1} \bar{1}  1  1 \bar{1} \bar{1}  1  1],53:[ 1  1 \bar{1} \bar{1}  1  1 \bar{1} \bar{1}  1  1 \bar{1} \bar{1}  1  1 \bar{1} \bar{1}],54:
[ 1 \bar{1}  1 \bar{1}  1 \bar{1}  1 \bar{1}  1 \bar{1}  1 \bar{1}  1 \bar{1}  1 \bar{1}],\nonumber \\
&55:[ 1 \bar{1} \bar{1}  1 \bar{1}  1  1 \bar{1} \bar{1}  1  1 \bar{1}  1 \bar{1} \bar{1}  1],56:[ 1 \bar{1}  1 \bar{1} \bar{1}  1 \bar{1}  1 \bar{1}  1 \bar{1}  1  1 \bar{1}  1 \bar{1}],57:
[ 1 \bar{1} \bar{1}  1  1 \bar{1} \bar{1}  1 \bar{1}  1  1 \bar{1} \bar{1}  1  1 \bar{1}],\nonumber \\
&58:[ 1  1 \bar{1} \bar{1} \bar{1} \bar{1}  1  1 \bar{1} \bar{1}  1  1  1  1 \bar{1} \bar{1}],59:[ 1 \bar{1}  1 \bar{1}  1 \bar{1}  1 \bar{1} \bar{1}  1 \bar{1}  1 \bar{1}  1 \bar{1}  1],60:
[ 1 \bar{1} \bar{1}  1 \bar{1}  1  1 \bar{1}  1 \bar{1} \bar{1}  1 \bar{1}  1  1 \bar{1}],\nonumber \\
&61:[1 1 1 1 1 1 1 1 1 1 1 1 1 1 1 1],62:[ 1  1  1  1 \bar{1} \bar{1} \bar{1} \bar{1}  1  1  1  1 \bar{1} \bar{1} \bar{1} \bar{1}]\nonumber \\
&63:[ 1 \bar{1} \bar{1}  1  1 \bar{1} \bar{1}  1  1 \bar{1} \bar{1}  1  1 \bar{1} \bar{1}  1],64:[ 1  1 \bar{1} \bar{1} \bar{1} \bar{1}  1  1  1  1 \bar{1} \bar{1} \bar{1} \bar{1}  1  1],\nonumber \\
%
&65:[ 1  1  1 \bar{1} \bar{1} \bar{1} \bar{1}  1  1 \bar{1} \bar{1} \bar{1} \bar{1}  1  1  1],66:[ 1 \bar{1}  1  1 \bar{1}  1 \bar{1} \bar{1}  1  1 \bar{1}  1 \bar{1} \bar{1}  1 \bar{1}],67:
[ 1 \bar{1} \bar{1} \bar{1} \bar{1}  1  1  1 \bar{1} \bar{1} \bar{1}  1  1  1  1 \bar{1}],\nonumber \\
&68:[ 1  1 \bar{1}  1  1  1 \bar{1}  1 \bar{1}  1 \bar{1} \bar{1} \bar{1}  1 \bar{1} \bar{1}],69:[ 1  1 \bar{1}  1 \bar{1} \bar{1}  1 \bar{1} \bar{1}  1 \bar{1} \bar{1}  1 \bar{1}  1  1],70:
[ 1 \bar{1} \bar{1} \bar{1}  1 \bar{1} \bar{1} \bar{1}  1  1  1 \bar{1}  1  1  1 \bar{1}]\nonumber \\
&71:[ 1  1 \bar{1}  1 \bar{1} \bar{1}  1 \bar{1}  1 \bar{1}  1  1 \bar{1}  1 \bar{1} \bar{1}],72:[ 1  1 \bar{1}  1  1  1 \bar{1}  1  1 \bar{1}  1  1  1 \bar{1}  1  1],73:
[ 1 \bar{1} \bar{1} \bar{1} \bar{1}  1  1  1  1  1  1 \bar{1} \bar{1} \bar{1} \bar{1}  1]\nonumber \\
&74:[ 1 \bar{1}  1  1  1 \bar{1}  1  1  1  1 \bar{1}  1  1  1 \bar{1}  1],75:[ 1 \bar{1}  1  1 \bar{1}  1 \bar{1} \bar{1} \bar{1} \bar{1}  1 \bar{1}  1  1 \bar{1}  1],76:
[ 1  1  1 \bar{1}  1  1  1 \bar{1} \bar{1}  1  1  1 \bar{1}  1  1  1]\nonumber \\
&77:[ 1 \bar{1} \bar{1} \bar{1}  1 \bar{1} \bar{1} \bar{1} \bar{1} \bar{1} \bar{1}  1 \bar{1} \bar{1} \bar{1}  1],78:[ 1  1  1 \bar{1}  1  1  1 \bar{1}  1 \bar{1} \bar{1} \bar{1}  1 \bar{1} \bar{1} \bar{1}],\nonumber \\
&79:[ 1 \bar{1}  1  1  1 \bar{1}  1  1 \bar{1} \bar{1}  1 \bar{1} \bar{1} \bar{1}  1 \bar{1}],80:[ 1  1  1 \bar{1} \bar{1} \bar{1} \bar{1}  1 \bar{1}  1  1  1  1 \bar{1} \bar{1} \bar{1}].\nonumber \\
\label{80rays}
\end{eqnarray}

In  (\ref{80rays}), each ray is paired with a partner ray (possibly itself), which is obtained by inversion of the entries in the original ray. The concept of a partner ray allows us to convert a  BKS proof (about contextuality) into a proof of Bell's theorem (about non-locality), as described in \cite{Aravind1999}.

In \cite{Harvey2008}, a non-parity BKS proof $80-265$ was proposed. Here we find a smaller $80-21$ one. Our strategy is as follows. Let us consider the set

$$D=\{\frac{1}{\sqrt{5}},\frac{\sqrt{3}}{\sqrt{10}},\frac{\sqrt{2}}{\sqrt{5}},\frac{1}{\sqrt{2}},\frac{\sqrt{3}}{\sqrt{5}},\frac{\sqrt{7}}{\sqrt{10}},\frac{2}{\sqrt{5}}\}\approx \{0.447,0.547,0.632,0.707,0.774,0.836,0.894\},$$
that characterizes the allowed distances between the $625$ bases. We randomly select a minimal set $B$ of $l$ bases within the $625$'s such that
(a) there is at least one distance of each type among the selected bases,
(b) there is at least one subset of $B$ containing $5$ bases partitioning the $5\times16=80$ rays (this criterion is adopted to reach the result with only $16^5=1048576$ checks),
(c) the set $B$ satisfies the BKS postulates (i) and (ii) listed in the introduction.

We found a mimimal cardinality $l=23$ for the set $B$. It was further simplified to $l=22$, a set still satisfying the criterion (b), then to $l=21$. The $80-21$ proof, given in (\ref{21bases}), does not satisfy criterion (b), although there exist two sets of four disjoint bases. The main properties of $80-23$, $80-22$ and $80-21$ proofs are summarized in Tables 5 and 6.

\begin{table}[ht]
\begin{center}
\small
\begin{tabular}{|r||r|r|r|r|r|r|r|}
\hline
 proof $v-l$ & $a_1$ & $a_2$ &  $a_3$ &$a_4$ & $a_5$ &  $a_6$ &$a_7$\\
\hline
$80-23$ & $1$ & $3$ &  $17$ &$19$ & $76$ &  $69$ &$68$\\
\hline
$80-22$ & $1$ & $1$ &  $17$ &$19$ & $65$ &  $64$ &$64$\\
\hline
$80-21$ & $1$ & $1$ &  $14$ &$19$ & $60$ &  $64$ &$51$\\
\hline
\end{tabular}
\label{table5}
\caption{The histogram of distances for various  proofs obtained from the square of operators (\ref{Mermin3}).}
\end{center}
\end{table}

\begin{table}[ht]
\begin{center}
\small
\begin{tabular}{|r||r|r|r|r|r|r|r|}
\hline
 \# common elements & $0$ &  $2$ & $4$ &  $6$  & $8$ & $10$ &  $12$  \\
 \hline
 distance & $a_7$ & $a_6$ & $a_5$ & $a_4$ & $a3$ & $a_2$ & $a_1$\\
\hline
\hline
$80-23$   & $\mathbb{Z}_1$& $D_6$ &   $\mathbb{Z}_2$ & $\mathbb{Z}_2^2\rtimes(A_7 \rtimes \mathbb{Z}_2)$ &   $\mathbb{Z}_2^5 \rtimes \mathbb{Z}_6$ &  $\mathbb{Z}_2^5 \times S_{18}$ & $\mathbb{Z}_2 \times S_{21}$ \\
\hline
$80-22$  & $\mathbb{Z}_1$&   $D_6$ &   $\mathbb{Z}_2$ & $\mathbb{Z}_2^2 \rtimes S_6$ &   $\mathbb{Z}_2^3 \rtimes S_6$ &  $\mathbb{Z}_2 \times S_{20}$ &  $\mathbb{Z}_2 \times S_{20}$ \\
\hline
$80-21$  &$\mathbb{Z}_1$ &    $\mathbb{Z}_2^2 $ &   $\mathbb{Z}_2$ & $\mathbb{Z}_2^2 \rtimes S_5$ &   $\mathbb{Z}_2^3 \rtimes S_5$ &  $\mathbb{Z}_2 \times S_{19}$ &  $\mathbb{Z}_2 \times S_{19}$ \\
\hline
\end{tabular}
\label{table6}
\caption{The symmetries involved in the selected four-qubit proofs of the BKS theorem. The first row gives the number of common elements between the bases, the second row providing the corresponding distances. The proof $80-23$ contains the $80-22$ one, and the latter contains the $80-21$ one. Thus Table 6 has a slightly different status than Tables 2 and 4, where only critical proofs were displayed. }
\end{center}
\end{table}

\noindent

\begin{eqnarray}
&1:   \{ 35, 37, 43, 45, 51, 53, 54, 57, 65, 69, 71, 72, 74, 76, 77, 80 \},\nonumber\\
&2:   \{ 17, 18, 25, 26, 35, 37, 39, 40, 43, 45, 47, 48, 51, 52, 57, 59 \},\nonumber\\
&3:   \{ 35, 40, 43, 48, 50, 52, 58, 59, 61, 62, 63, 64, 68, 70, 78, 79 \},\nonumber\\
&4:    \{ 1, 2, 4, 5, 7, 11, 12, 16, 21, 22, 25, 26, 35, 37, 39, 48 \},\nonumber\\
&5:    \{ 3, 7, 8, 16, 18, 19, 20, 21, 24, 25, 30, 31, 33, 42, 44, 46 \},\nonumber\\
&6:    \{ 1, 2, 3, 4, 6, 7, 8, 9, 10, 12, 14, 16, 19, 20, 24, 31 \},\nonumber\\
&7:       \{ 1, 6, 10, 12, 23, 24, 31, 32, 33, 34, 36, 46, 49, 60, 62, 64 \},\nonumber\\
&8:    \{ 17, 18, 20, 21, 22, 25, 26, 27, 29, 30, 31, 32, 37, 43, 47, 48 \},\nonumber\\
&9:    \{ 20, 27, 31, 32, 33, 34, 36, 37, 38, 41, 42, 43, 44, 46, 47, 48 \},\nonumber\\
&10:   \{ 1, 2, 9, 10, 20, 27, 31, 32, 37, 43, 47, 48, 52, 53, 57, 63 \},\nonumber\\
&11:    \{ 3, 7, 11, 15, 33, 34, 41, 42, 54, 55, 57, 58, 59, 60, 63, 64 \},\nonumber\\
&12:    \{ 1, 2, 3, 4, 5, 6, 7, 8, 9, 10, 11, 12, 13, 14, 15, 16 \},\nonumber\\
&13:    \{ 2, 3, 10, 11, 12, 13, 14, 16, 65, 66, 74, 75, 76, 78, 79, 80 \},\nonumber\\
&14:    \{ 33, 36, 41, 44, 56, 58, 60, 62, 65, 69, 70, 73, 74, 75, 77, 79 \},\nonumber\\
&15:    \{ 4, 6, 12, 14, 51, 54, 56, 58, 59, 60, 61, 62, 65, 69, 73, 75 \},\nonumber\\
&16:    \{ 18, 21, 26, 29, 49, 50, 55, 64, 66, 67, 68, 71, 76, 77, 79, 80 \},\nonumber\\
&17:    \{ 5, 11, 13, 15, 21, 22, 23, 27, 28, 29, 30, 32, 53, 54, 61, 63 \},\nonumber\\
&18:    \{ 5, 11, 13, 15, 17, 18, 23, 25, 26, 27, 28, 32, 51, 52, 57, 59 \},\nonumber\\
&19:   \{ 33, 34, 36, 38, 41, 42, 44, 46, 65, 66, 67, 69, 71, 73, 75, 80 \},\nonumber\\
&20:    \{ 17, 18, 20, 21, 22, 24, 25, 26, 28, 29, 30, 32, 67, 69, 75, 80 \},\nonumber\\
&21:    \{ 1, 6, 10, 12, 33, 34, 36, 39, 40, 46, 47, 48, 66, 67, 73, 75 \}.\nonumber\\
\label{21bases}
\end{eqnarray}

For the sake of completeness, we mention that the $22$-base and $23$-base proofs follow by adding to (\ref{21bases}) the following two rays, respectively


\begin{eqnarray}
&\{ 2, 10, 12, 14, 35, 37, 43, 45, 65, 69, 71, 74, 76, 78, 79, 80 \}, \nonumber \\
&\{ 49, 50, 55, 56, 58, 60, 62, 64, 68, 70, 72, 74, 76, 77, 78, 79\}.  \nonumber \\
\end{eqnarray}

That (\ref{21bases}) is a BKS proof of the four-qubit system
can be easily checked with the help of a computer by checking that for all $16^4*80=5242880$ possibilities of assigning the truth value $1$ to a quintuple of rays $(i,j,k,l,m)$ with $i$, $j$, $k$, $l$ and $m$ being the indices
in one set of four mutually disjoint bases and in an arbitrary base of index $m$ of (\ref{21bases}), at least one basis does not satisfy the constraint (ii) of the introduction. The same conclusion holds for the set of $22$ bases that contains the set of the $21$'s, and for the set of $23$ bases that contains the set of the $22$'s. No further simplification of the $21$-base set could be obtained while keeping the BKS proof.

One observes from Table 6 (column 2) that the proofs are quite random given the overall symmetry group $\mathbb{Z}_1$. But many remnant symmetries are present as one can see by looking at the other crossing graphs (in columns 3 to 8 ).

\section{Conclusion}

We have performed a systematic investigation of small state proofs of the BKS theorem involving real rays of several qubits. The proofs correspond to some sets of maximal orthogonal bases constructed from Mermin's $3\times 3$ square (for two qubits) and from Mermin's pentagram (for three and four qubits). These BKS states belong to a larger set of real states on an (Euclidean) Barnes-Wall lattice $B_n$. It would be desirable to discover the precise status of the KS sets on $B_n$. This is left for a future work.

Another ongoing work of ours concerns BKS proofs with complex rays in the spirit of \cite{Aravind2011new, Megill2011} and BKS proofs for more qubits (a particular case of five qubits is investigated in \cite{PlaSan5QB}). A deeper understanding of KS sets may be useful for conceptual questions concerning the EPR local elements of reality, quantum complementarity, counterfactual compatibility and non-contextual inequalities \cite{Liang2011,Abramsky2012}.

\section*{Acknowledgements} The author thanks the anonymous referees for their constructive remarks. He also acknowledges Metod Saniga for his help in improving the second version of the manuscript.

\end{document}